\documentstyle[emulateapj]{article}

\lefthead{Armitage}
\righthead{Turbulence in accretion disks}

\begin{document}

\submitted{ApJ Letters, in press}

\title{Turbulence and angular momentum transport in a global 
       accretion disk simulation}

\author{Philip J. Armitage}
\affil{Canadian Institute for Theoretical Astrophysics, McLennan Labs,
	60 St George St, Toronto, M5S 3H8, Canada}

\begin{abstract}
The global development of magnetohydrodynamic turbulence in an accretion
disk is studied within a simplified disk model that omits vertical
stratification. Starting with a weak vertical seed field, a
saturated state is obtained after a few tens of orbits in which the 
energy in the predominantly toroidal magnetic field is still subthermal.
The efficiency of angular momentum transport, parameterized by the
Shakura-Sunyaev $\alpha$ parameter, is of the order of $10^{-1}$.
The dominant contribution to $\alpha$ comes from magnetic stresses, 
which are enhanced by the presence of weak net vertical fields. The
power spectra of the magnetic fields are flat or decline only slowly
towards the largest scales accessible in the calculation, suggesting
that the viscosity arising from MHD turbulence may not be a locally
determined quantity. I discuss how these results compare with 
observationally inferred values of $\alpha$, and possible 
implications for models of jet formation.
\end{abstract}	


{\em For visualization of simulations see
{\tt http://www.cita.utoronto.ca/$^\sim$armitage/global{\underline{ }}abs.html}}

\section{INTRODUCTION}

The Balbus-Hawley instability is the most generally applicable
mechanism known to initiate turbulence and outward angular momentum
transport in accretion disks (Balbus \& Hawley 1991). This is a linear, 
local instability that exists for rotating flows threaded by a weak 
magnetic field with ${\rm d} \Omega^2 / {\rm d} r < 0$, conditions satisfied in 
disks (for earlier discussions see Velikhov 1959, Chandrasekhar 1961). 
A vigorous growth rate is obtained for
a wide variety of initial magnetic field configurations (Balbus \& 
Hawley 1992; Ogilvie \& Pringle 1996; Terquem \& Papaloizou 1996), 
implying that the instability is inescapable for ionized disks
where the field is well-coupled to the gas.

Extensive numerical simulations have explored the nonlinear 
development of the instability within the local, shearing box
approximation (for a review, see e.g. Gammie 1998). Such
simulations have convincingly established that the nonlinear
development of the Balbus-Hawley instability leads to sustained
turbulence and significant angular momentum transport, typically
finding a Shakura-Sunyaev (1973) $\alpha \approx 10^{-2}$ 
(Hawley, Gammie \& Balbus 1995, 1996; Stone et al 1996; 
Brandenburg et al. 1995). There is some evidence for 
cyclic behavior that might have important implications 
for disk variability (Brandenburg et al. 1996).
Equally important has been the final elimination of convection (Stone \& 
Balbus 1996), and the near-elimination of nonlinear hydrodynamic 
turbulence (Balbus, Hawley \& Stone 1996), as plausible rival
mechanisms for angular momentum transport in accretion disks.
Progress has also been made in trying to understand how the rich
phenomenology of accretion disk variability can arise within a
dynamo driven disk model (Armitage, Livio \& Pringle 1996;  
Gammie \& Menou 1998), although much more remains to be done 
in this area.

There are many further questions that one may hope simulations 
will address, and not all of them are amenable to a local 
treatment. Most obviously, is the angular momentum transport
in a disk locally determined? What is the structure of the 
spatial and time variability of the disk fields, and are they
suitable for launching a magnetically driven disk wind or jet?
Unsurprisingly, the global calculations needed to investigate
these issues are extremely demanding, both as a consequence
of the larger computational domain and, especially, because
of the need to simulate regions of low density where the 
high Alfv\'en speed severely limits the timestep of explicit
numerical codes.

In this paper, results are presented from a vertically unstratified
global simulation of accretion disk turbulence. Such a calculation
is evidently missing essential physics. There is no buoyancy, 
no possibility of Parker instability (Parker 1979), and no 
magnetically dominated disk corona -- all features that are expected 
to arise in a full disk model and which may be crucial for the 
disk dynamo problem (Tout \& Pringle 1992). However the lesser computational
demands permit a preliminary investigation of some of the
important questions raised by previous, local, simulations.

\section{NUMERICAL SIMULATION}

The equations of ideal magnetohydrodynamics (MHD) are solved using the
ZEUS-3D code developed by the Laboratory for Computational Astrophysics
(Clarke, Norman \& Fiedler 1994; Stone \& Norman 1992a, 1992b). ZEUS is
a time explicit eulerian finite difference code that uses the method of 
characteristics (MoC) -- constrained transport scheme to evolve the magnetic
fields (Hawley \& Stone 1995; Stone \& Norman 1992b). For this 
simulation an isothermal equation of state $P = \rho c_s^2$ replaces 
the internal energy equation, so the remaining equations are:
\begin{eqnarray}
 { {\partial \rho} \over {\partial t} } + \nabla \cdot
 \left( \rho \vec{v} \right) & = & 0 \\
 { {\partial \left( \rho \vec{v} \right)} \over {\partial t} } + 
 \nabla \cdot \left( \rho \vec{v} \vec{v} \right) & = &  
 - \nabla P - \rho \nabla \Phi + \vec{J} \times \vec{B} \\
 { {\partial \vec{B}} \over {\partial t} } & = & 
 \nabla \times \left( \vec{v} \times \vec{B} \right),
\end{eqnarray} 
where the symbols have their conventional meanings. There is no
explicit resistivity in the calculation, reconnection occurs 
numerically on the grid scale. We use second order interpolation
(van Leer 1977) for all advected quantities, and the latest 
(and allegedly most stable) version of the MoC algorithm for 
the induction equation.

\subsection{Initial and boundary conditions}

The calculation is performed in cylindrical polar geometry
$(r, \phi, z)$, in a volume bounded by $r_{\rm in} = 1$, 
$r_{\rm out} = 4$, and $z = \pm H = 0.4$. We take 
$\Phi = \Phi(r) = 1/r$, implying no vertical component of
gravity. The boundary conditions are periodic in $z$, reflecting at
$r = r_{\rm in}$ ($v_r = B_r = 0$), and set to outflow at $r=r_{\rm out}$.
Outflow boundary conditions are implemented as a
simple extrapolation of fluid variables on the grid into
the boundary zones. These boundary conditions admit the 
development of net vertical flux through the computational 
volume as material leaves the grid -- the import of this 
for the resulting $\alpha$ will be discussed later.
We use a grid of $(n_r=128, n_\phi=320, 
n_z=32)$ zones, with uniform zoning.

An initial state is obtained by evolving a disk with constant
surface density and Keplerian rotation in two dimensions until 
transients due to pressure gradients and the influence
of the inner boundary condition die out. This requires a time
$\Delta t=100$, where time is measured in units of the 
orbital period at the inner edge. The resulting equilibrium state has
$\Sigma = {\rm constant}$ between $R=1.4$ and $R=3$, tapering to 
a low value at the boundaries. The velocity profile is Keplerian 
to better than 10\% over the entire radial range. No evidence is
found for purely hydrodynamic disk instabilities. We then add a weak 
vertical magnetic field that is non-zero only between $r=1.5$ and $r=3.5$,
of the form,
\begin{equation}
 B_z (r) = { B_0 \over r } \sin \left( \pi \left( r - 1.5 \right) \right),
\end{equation} 
chosen simply to be divergence free and to have zero net flux in 
the $z$ direction. The simulation is then evolved for a further 80 
orbits. The computational cost of this using ZEUS is approximately
$2 \times 10^{14}$ floating point operations -- a large but not 
outrageous number for current workstations.

\subsection{Results}

The initial state is immediately unstable, leading to rapid
growth in the magnetic and perturbed velocity fields. Around
40 inner orbits of evolution were required before a saturated
state was obtained, which was then followed for an additional
time $\Delta t = 40$ without any further qualitative changes in 
disk behavior occurring. The magnetic energy in the saturated
state is dominated by the toroidal field component and is
subthermal, $\sim 0.2-0.3$ of the thermal energy, the corresponding
energies in $B_r$ and $B_z$ are respectively $10^{-2}$ and 
$6 \times 10^{-3}$ of the thermal energy. The velocity 
perturbations are mildly anisotropic, $v_z$ and $\delta v_\phi$
display roughly gaussian distributions with width $\approx 0.3 c_s$, 
while that of $v_r$ has a width of $\approx 0.4-0.5 c_s$. The 
total energy in the perturbed velocity field is an order of 
magnitude below that of the magnetic fields.

Fig.~1 shows the surface density and vertically averaged 
$B_z$ component of the magnetic field at the
conclusion of the calculation. We
plot the overdensity $\delta \Sigma$ relative to the
azimuthally averaged surface density profile,
\begin{equation}
 \delta \Sigma (r,\phi) = { {2 \pi \Sigma (r,\phi)} \over
 {\int_0^{2 \pi} \Sigma (r,\phi) d \phi} }.
\end{equation}
The magnetic field is plotted without any such normalization.

The combination of turbulence and shear leads to a ragged spiral pattern 
in the surface density, though the azimuthal fluctuations are 
relatively small and described by a gaussian with a width 
$\sim 0.35 \Sigma(r)$. The magnetic field displays 
a filamentary structure, both in projection and when visualized 
in three dimensions. Visually it is evident that the field 
shows considerable coherence over large scales, especially in 
azimuth. $B_r$ and $B_\phi$ (not plotted) show similar patterns.

\subsection{Efficiency of angular momentum transport}

The efficiency of angular momentum transport in the simulation
can be measured by $\alpha$, the ratio of the shear stress 
$w_{r \phi}$ to the gas pressure $P$ in the form,
\begin{equation} 
 \alpha = {2 \over 3} {w_{r \phi} \over P}.
\end{equation} 
This is consistent with a relation between the shear viscosity 
$\nu$ and the disk sound speed $c_s$ of the usual form, $\nu = \alpha 
c_s^2 / \Omega$, with $\Omega$ the Keplerian angular velocity.
The magnetic and fluid contributions to the total stress
are then given by (e.g. Gammie 1998),
\begin{equation} 
 \alpha_{M}  =  {2 \over 3} \left\langle { {-B_r B_\phi} \over
 {4 \pi P} } \right\rangle , \ \ \
 \alpha_{R}  =  {2 \over 3} \left\langle { {\rho v_r \delta v_\phi} 
 \over {P} } \right\rangle , 
\end{equation} 
respectively, where the brackets denote an average in the spatial 
co-ordinates. 

Fig.~2 shows the time evolution of $\alpha_{M}$ and $\alpha_{R}$ 
at a single radius $r = 0.5 (r_{\rm in} + r_{\rm out})$ in the 
disk. Apart from some transient waves at early times, $\alpha$ from 
both fluid and magnetic stresses is positive, with the bulk of the 
angular momentum transport being provided by magnetic stresses. 
At the end of the calculation, the total $\alpha$ {\em at this 
radial location} is in the range $\alpha = 0.05 - 0.1$.

Evaluating $\alpha_{M}$ and $\alpha_{R}$ as functions of radius
at $t=80$, $\alpha_{R}$ is found to be roughly constant across 
the grid in the range $\alpha_{R} = 0.01 - 0.02$. $\alpha_{M}$ 
varies by factors of a few, and is close to its minimum value 
at the location in the center of the grid used for plotting 
the time series in Fig.~2. Taking an average over the entire
simulation volume, we obtain,
\begin{equation} 
 \alpha_{M} \simeq 0.17, \ \ \ 
 \alpha_{R} \simeq 2.0 \times 10^{-2}. 
\end{equation}
The value of $\alpha$ can also be estimated by comparing the
simulation to the results of one dimensional diffusive disk
models (e.g. Pringle 1981). Comparing the evolution of the radial center 
of mass between the two calculations one again obtains an $\alpha$ 
value of $\sim 0.1$, though as mentioned already $\alpha$ is 
unsurprisingly not a constant in the simulation.

\subsection{Influence of net vertical magnetic field}

The choice of outflow boundary conditions at $r_{\rm out}$ was 
motivated by the desire to reduce the amplitude of reflected 
waves at the edge of the grid. However these boundary conditions
also allow the development of net vertical magnetic fields that 
can have a strong influence on the properties of disk turbulence.
Previous work has found that a net field significantly enhances 
the strength of turbulence and boosts the derived value of 
$\alpha$ (Hawley, Gammie \& Balbus 1995). Adopting the parameterization 
of the results of local simulations given by Gammie (1998),
\begin{equation} 
 \alpha \sim 0.01 + 4 { { \langle v_{Az} \rangle } \over c_s } 
 + {1 \over 4} { { \langle v_{A \phi} \rangle } \over c_s }, 
\end{equation}
where $v_{Az}$ is the Alfven velocity corresponding to the net
vertical field, we find that the enhancement of $\alpha$ in 
this simulation due to net fields is expected to be $\sim 0.05$. 
This is obviously a very crude estimate, as we are using the 
above relation in an untested regime, but it does imply that 
it is premature to conclude that global simulations lead to 
higher values of $\alpha$ than local calculations.

\subsection{Coherence of the magnetic field}

The scale of the magnetic fields generated in the disk is
analyzed via the azimuthal Fourier decomposition,
\begin{equation}
 C_{m,i} (r,z) = { 1 \over {2 \pi} } \int_0^{2 \pi} B_i e^{-i m \phi} d \phi.
\end{equation} 
Fig.~3 plots $m \vert C_m \vert^2$ for the three field components, 
in each case averaged over the the entire volume of the simulation.

The azimuthal power spectra show similar shapes for each of the 
magnetic field components. There is a power-law decline somewhat 
steeper than that expected from Kolmogorov turbulence at 
large $m$, and a break at $m \simeq 10-20$, corresponding to a
physical scale of $\sim H$ in the center of the grid. 
This confirms the visual impression that the magnetic fields 
are patchy with typical scales of the order of $H$.
However there is also considerable power at the largest scales, 
with the power per logarithmic interval in $m$ declining only 
slowly towards low $m$ for all field components in the range $m=1-10$. 
Similar conclusions follow from the radial power spectra.

\section{DISCUSSION}

In this paper, we have reported on a global simulation of 
an unstratified magnetized accretion disk. As expected from 
analytic considerations (Curry \& Pudritz 1994, 1995, 1996) and 
local simulations, MHD instabilities generate sustained turbulence 
that leads to outward transport of angular momentum. The 
generated fields possess considerable power in azimuthal 
modes of low $m$, which correspond to physical scales 
considerably in excess of $H$, the disk semi-thickness,
and display a ragged spiral structure. The current simulation
does not admit the development of the Parker instability, which
might depress the power on large scales, but with this caveat 
the results suggest that a viscosity originating from MHD
turbulence may not be a locally determined quantity. The dominant 
magnetic field component is toroidal, and the interaction of this 
fluctuating internal field with a magnetosphere is likely 
to be an important complication to the already complex picture 
of star-disk interaction in magnetic systems (Miller \& Stone 1997;
Torkelsson 1998).  

The efficiency of angular momentum transport seen in the
calculation, parameterized by the Shakura-Sunyaev $\alpha$ 
prescription, is $\alpha \approx 10^{-1}$. This is larger
than the value obtained from local calculations (Gammie 1998) 
with zero net vertical magnetic fields, though we have noted
that the influence of a vertical field on the current simulation
is likely to have boosted the value of $\alpha$ significantly.
More realistic simulations with demonstrated numerical convergence 
are evidently required. However there is no obvious discrepancy 
with the values of $\alpha$ inferred for dwarf novae, where modeling 
of disk outbursts suggests $\alpha = 0.1 - 0.3$ (Cannizzo 1993),
and for Active Galactic Nuclei, where the admittedly poorer observations are 
consistent with an $\alpha$ of $10^{-2}$ (Siemiginowska \& 
Czerny 1989). Conversely it is hard to see why a viscosity 
derived from MHD turbulence should be two or three orders
of magnitude {\em lower} in the ionized inner regions of protostellar
disks, as required to match the timescales of FU Orionis
outbursts within disk instability models (Bell \& Lin 1994).
This may call into question the {\em self-regulated} aspect of the thermal 
disk instability picture for FU Orionis events. The generation of magnetic 
fields of large scale may additionally be important for 
models of jet formation (eg. Blandford \& Payne 1982; Ouyed, Pudritz 
\& Stone 1997; Matsumoto \& Shibata 1997; Konigl 1997), which if 
generated via a disk dynamo would be expected to be most efficient 
in relatively thick disks or advection dominated flows (Narayan \& 
Yi 1995). Observations of which systems produce jets appear to be
broadly consistent with a model in which $(H/R)$ is a controlling 
parameter, though many other possibilities are also viable (Livio 1997).

\acknowledgements

I thank Jim Stone and James Murray for valuable discussions at the 
start of this work, and the referee for a very 
prompt and helpful report. I am grateful to Mark Bartelt for 
ensuring the availability of the required computing resources.

\newpage            

\begin{figure}
\plotone{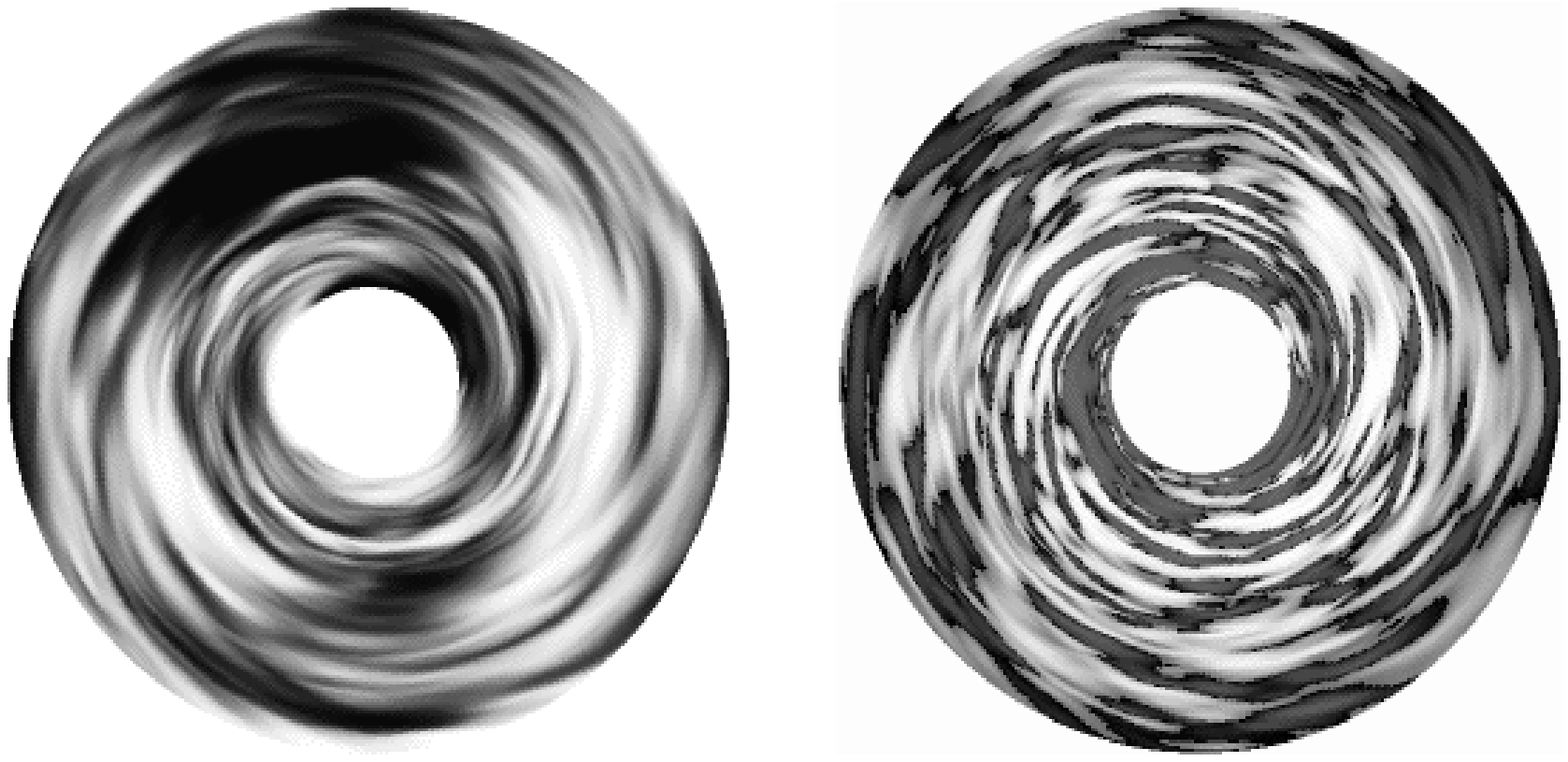}
\caption{Maps of the disk surface overdensity (left image) and 
            integrated vertical component of the magnetic field, 
            $B_z$ (right image), at $t=80$ orbits. The surface
            density is plotted relative to the azimuthally 
            averaged value, ie $\Sigma(r,\phi) / \Sigma(r)$.
            Typical azimuthal fluctuations in $\Sigma$ are at the 
            tens of percent level, typical $B_z$ fields 
            are of the order of $10^{-3}$ of the thermal 
            energy. The disk rotates clockwise.}
\end{figure}            
            
\begin{figure}
\plotone{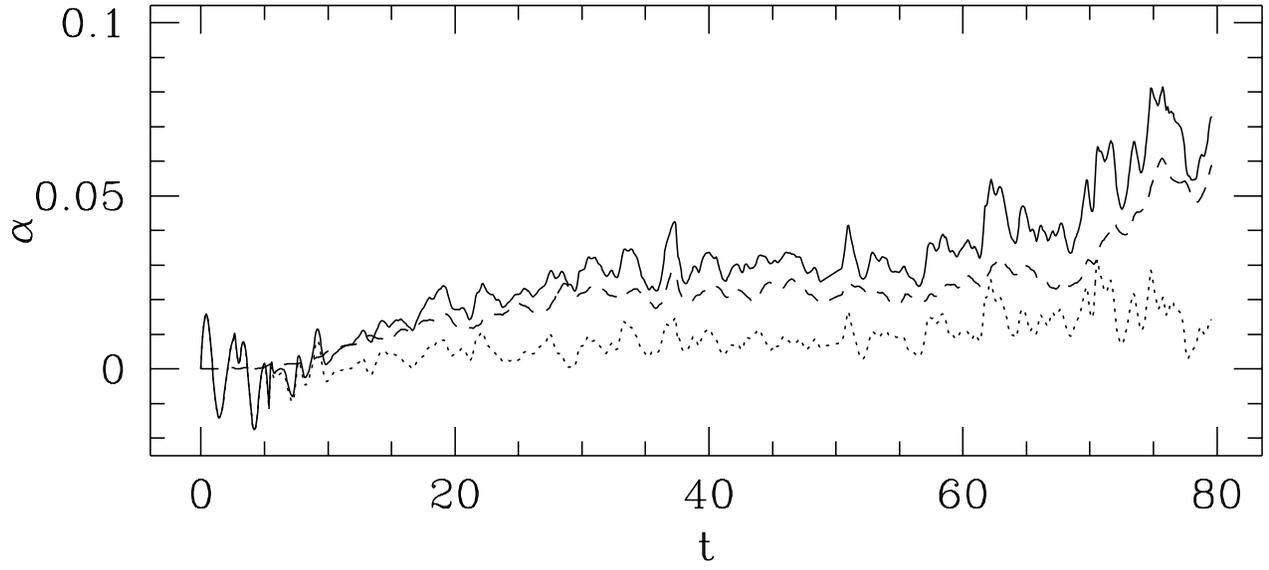}            
\caption{Derived Shakura-Sunyaev $\alpha$ parameter at the center 
            of the grid. The solid line shows the total $\alpha$, the 
            dashed and dotted lines the contribution from magnetic 
            and fluid stresses respectively.}
\end{figure}
            
\begin{figure}
\plotone{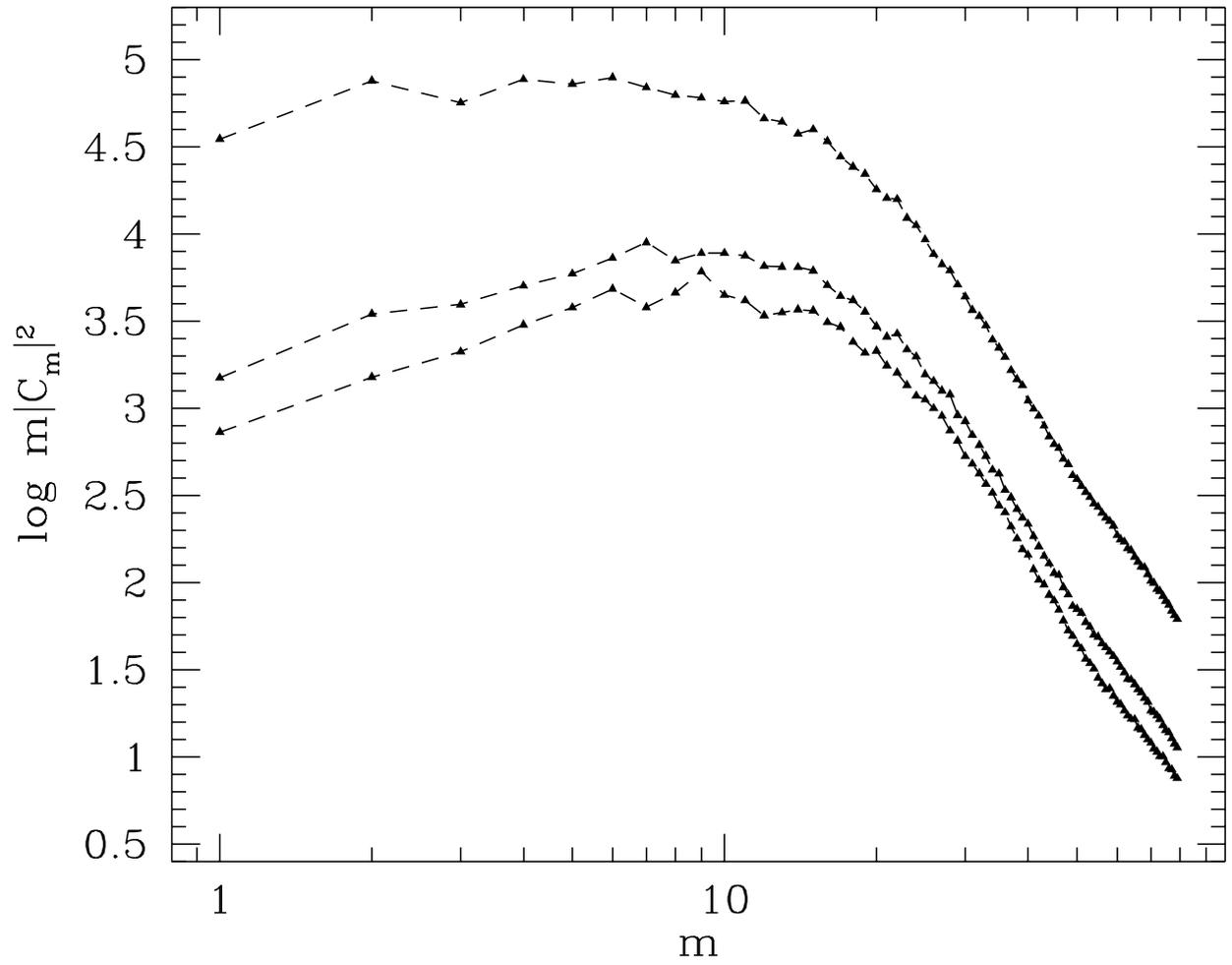}
\caption{Fourier decomposition of the azimuthal structure of 
            the disk magnetic field, averaged over the simulation
            volume. From top downwards, the curves show the 
            power spectra for $B_\phi$, $B_r$ and $B_z$ respectively.}
\end{figure}     

\end{document}